\documentclass[aps,superscriptaddress,twocolumn,a4paper,10pt]{revtex4-2}
\usepackage[colorlinks=true, pdfstartview=FitV, linkcolor=blue, citecolor=red, urlcolor=black]{hyperref}
\usepackage{graphicx}
\usepackage[all]{xy}
\usepackage{amsmath}
\usepackage{amssymb}
\usepackage{tensor}
\usepackage{orcidlink}
\usepackage{amsthm}
\usepackage{comment}
\newcommand{\be}{\begin{equation}}
\newcommand{\ee}{\end{equation}}
\newcommand{\ben}{\begin{eqnarray}}
\newcommand{\een}{\end{eqnarray}}
\newcommand{\bes}{\begin{subequations}}
\newcommand{\ees}{\end{subequations}}
\def\bal#1\eal{\begin{align}#1\end{align}}

\newcommand{\bfi}{\begin{figure}}
\newcommand{\efi}{\end{figure}}
\newcommand{\bc}{\begin{center}}
\newcommand{\ec}{\end{center}}

\newcommand{\sech}{{\rm sech}}
\newcommand{\arctanh}{{\rm arctanh}}
\newcommand{\p}{{\partial}}
\newcommand{\LL}{{\cal L}}

\begin{document}

\title{Analytical short- and long-range kink-like structures\\
in scalar field models with polynomial interactions}
%-------------------------------------------%
\author{I. Andrade\,\orcidlink{0000-0002-9790-684X}}
        \email[]{andradesigor0@gmail.com}\affiliation{Departamento de F\'\i sica, Universidade Federal da Para\'\i ba, 58051-970 Jo\~ao Pessoa, PB, Brazil}
\author{M.A. Marques\,\orcidlink{0000-0001-7022-5502}}
        \email[]{marques@cbiotec.ufpb.br}\affiliation{Departamento de Biotecnologia, Universidade Federal da Para\'\i ba, 58051-900 Jo\~ao Pessoa, PB, Brazil}
\author{R. Menezes\,\orcidlink{0000-0002-9586-4308}}
        \email[]{rmenezes@dcx.ufpb.br}\affiliation{Departamento de Ci\^encias Exatas, Universidade Federal
da Para\'{\i}ba, 58297-000 Rio Tinto, PB, Brazil}\affiliation{Departamento de F\'{\i}sica, Universidade Federal de Campina Grande,  58109-970 Campina Grande, PB, Brazil}

\begin{abstract}
We investigate a class of scalar field models which engender kink-like solutions in the presence of polynomial potentials that allows for modifications of the tails of the localized configurations. We introduce a parameter in the potential that controls the classical mass associated to its minima. By using the first-order framework developed by Bogomol'nyi, we obtain analytical solutions that become more and more interactive as we increase such parameter. By investigating the limit in which the parameter tends to infinite, the kink solution gets power law tails, and we show that this feature is related to the behavior of the classical mass, which vanishes in the aforementioned limit. We also investigate the stability against small fluctuations, with the results unveiling that, depending on the values of the parameter, the stability potential may support several bound states and also, it may attain a volcano-like profile.

\end{abstract} 

%\pacs{11.27.+d}
\maketitle

\section{Introduction}
Kinks are topological structures that arise under the action of a single real scalar field in $(1,1)$ dimensions \cite{vachaspati}. They are static solutions of the equation of motion which connect two distinct minima of the potential. A well-known model which supports these structures is the so-called $\phi^4$ model. In this case, the kink is described by a hyperbolic tangent, so its asymptotic behavior is given by an exponential function. These objects are of current interest in Physics due to their applications in distinct areas of nonlinear science. For instance, they can be used to describe specific behaviors in magnetic materials \cite{fradkin}, to study two and three-body interaction in Bose-Einstein condensates \cite{bec1,bec2} and to model braneworld scenarios with a single extra dimension of infinite extent \cite{RS2,dewolfe}. Modifications in the internal structure of kinks are also of interest, since they can be used in geometrically constrained systems of magnetic domain walls \cite{constrained}. In this direction, one may consider the inclusion of a parameter in the potential, as proposed in Ref.~\cite{tdlee,BL}, which modifies the slope of the kink.

Another issue of current interest related to kinks is collision. This line of study started in Ref.~\cite{colisao0} and has been conducted in many papers over the years; see, e.g., Refs.~\cite{colisao1,colisao2,colisao3,colisao4,colisao5,colisao6} and references therein. In this context, the tails of the kinks to be collided are very important, as they determine how the structures will interact. Therefore, solutions with non-exponential falloff may lead to different results. By analyzing canonical scalar field models, i.e., models in which the Lagrangian density is the difference between kinetic and potential terms, one can show that the asymptotic behavior of the static solutions is associated to the classical masses of the potential at the minima connected by the field. Therefore, to get topological structures with non-exponential tails, one must consider potentials with other classical masses.

Potentials supporting minima with infinite classical masses were proposed in Refs.~\cite{compact1,compact2,compact3,flores}. In this situation, the solution may become compact, i.e., it may attain its boundary values in a compact space. In this case we may say that the solution engenders a short-range behavior. In Refs.~\cite{kinktocompacton1,kinktocompacton2}, it was shown how to compactify the kink, making the falloff become faster as the classical mass increases. This means that the solution becomes less and less interactive as the compactification process occurs. This was also investigated in the context of collisions in Ref.~\cite{colisaocomp}, which unveiled the appearance of metastable structures in the process. 

One may also go around and consider the opposite direction, making the kink become more interactive. This was considered in Refs.~\cite{long1,long2}. In particular, in Ref.~\cite{long2}, it was shown that potentials which support minima with null classical mass, described by an specific $\phi^6$ model, allows for the presence of solutions whose tails are given by power-law falloff, extending farther the exponential one. Due to this feature, these solutions are commonly called long-range or highly interactive. In the study of the linear stability, they may lead to strong resonance peaks. Several papers dealing with long-range structures have appeared in the literature; see Refs.~\cite{colisao4,long3,lohe,mantonlong,gani1,gani2,gani3,colisaolongcampos,colisaotranslong,colisaolongimp,oscillonsmass0}. In particular, in Ref.~\cite{mantonlong}, long-range kink-kink and kink-antikink forces that arise in octic potentials were investigated; the force decays with the fourth power of the distance between the objects. A generalization for higher order potentials, $\phi^{2n+4}$, was done in Ref.~\cite{gani1}, showing that the force decays with the $2n/(n-1)$-th power of the separation. In Ref.~\cite{colisaolongcampos}, the authors investigated interactions between long-range structures, showing that the annihilation of the long-range kink and antikink may occur directly into radiation, before forming a bion, constrasting with the case in which the kink engenders exponential tails.

Even though models supporting analytical solutions with highly interactive tails are well known, there is still no path to go from exponential to power-law tails described by analytical solutions in polynomial potentials. In Refs.~\cite{colisao4,sineduplobazeia}, it was introduced a parameter which controls the location of the minima in the double sine-Gordon potential. In this paper, we get inspiration from the models proposed in Refs.~\cite{tdlee,BL}, although we use the parameter to modify the minima of the potential instead of its maximum, to introduce a class of polynomial potentials which allows us to get analytical solutions that go from exponential to power-law tails, becoming more interactive as the parameter gets larger.   

\section{Model}
Our interest is to study kinks, so we consider the action $S$ in $(1+1)$ spacetime dimensions, with metric tensor $\eta_{\mu\nu}=\text{diag}(+,-)$. The action is $S=\int dx\,dt \;\LL$, and the Lagrangian density has the canonical form
\be\label{lagrangian}
\LL = \frac12\p_\mu\phi\p^\mu\phi -V(\phi),
\ee
where $\phi$ is a real scalar field and $V(\phi)$ is the potential which specifies the model under investigation. We suppose that $V(\phi)$ is a non-negative function that engenders a set of minima, $v_i$, such that $V(v_i)=0$. We are considering dimensionless field and coordinates for simplicity. Associated to the minima of the potential, we can define the classical mass
\be\label{classicalmass}
m^2_{v_i} = V_{\phi\phi}\big|_{\phi=v_i},
\ee
with $V_{\phi\phi}$ representing the second derivative of the potential with respect to the scalar field. The equation of motion takes the form
\be\label{eom}
\Box\phi + V_\phi=0,
\ee
in which $\Box=\partial^2/\partial t^2 - \partial^2/\partial x^2$ is the d'Alambertian operator and $V_\phi$ is the derivative of the potential with respect to $\phi$. Since the Lagrangian density \eqref{lagrangian} is invariant under spacetime translations, one can obtain the conserved energy-momentum tensor $T_{\mu\nu} = \p_\mu\phi\p_\nu\phi -\eta_{\mu\nu}(\frac12\p_\alpha\phi\p^\alpha\phi -V(\phi))$ that leads to the following energy density
\be\label{rho}
\rho = \frac12\dot{\phi}^2 +\frac12{\phi'}^2 +V(\phi),
\ee
where dot and prime represent derivative with respect to time and spatial coordinates, respectively. As kinks are solutions of the equation of motion which minimize the energy of the system, we follow the method of Bogomol'nyi \cite{bogo}. If the potential is non negative, one may introduce an auxiliary function $W=W(\phi)$ to write $V(\phi)=(1/2)W_\phi^2$. This class of potentials support minimum energy configurations if the solutions are static, $\phi=\phi(x)$, and the following first-order equations are obeyed:
\be\label{fo}
\phi^\prime= \pm \sqrt{2V}.
\ee
The upper/lower sign denote the increasing/decreasing solution. Since both solutions are related by the change $x\to-x$, we only work with the equation with upper sign.

We want to modify the tail of static solutions. So, let us analyze its behavior for a general $V(\phi)$, by expanding the field around a minimum of the potential, $\phi(x)=v_i + \Tilde{\phi}(x)$. The terms of lowest orders in the equation of motion \eqref{eom} are
\be\label{fluc}
\Tilde{\phi}'' = m^2\Tilde{\phi} +\mu\Tilde{\phi}^\alpha,
\ee
where $m$ is the classical mass defined in Eq.~\eqref{classicalmass}, and $\mu$ and $\alpha>1$ are constants which depend on the model. If $m\neq0$, i.e., the term of lowest order is linear in the field, controlled by the classical mass, so the tail has an exponential falloff, $\Tilde{\phi}\propto e^{-m_{vi}|x|}$. This can be seen, for instance, in the well-known $\phi^4$ potential,
\be\label{vphi4}
V(\phi) = \frac12\left(1-\phi^2\right)^2,
\ee
which supports two minima, at $\phi=\pm1$, both with classical mass $m_\pm^2=4$. In this case, $\phi(x)=\tanh(x)$, so the solution has an exponential tail, $\Tilde{\phi}(x)\propto e^{-2|x|}$. There are several other potentials with this feature, including the sine-Gordon and polynomials of higher powers; see Refs.~\cite{tdlee,lohe,colisao4,bazeiapoly,ganipoly1,ganipoly2,perupoly1,ganipoly3}.

The case in which the classical mass is null ($m=0$) leads to a distinct tail. Since $\alpha>1$, Eq. \eqref{fluc} leads us to get $\Tilde{\phi}(x) \propto |x|^{-2/(\alpha-1)}$, so the falloff is slower, with power-law tails, and these configurations are called long-range solutions. This occurs, for example, in the potential
\be
V(\phi) = \frac12\left(v^2-\phi^2\right)^2\left(1-\phi^2\right)^2
\ee
considered in Refs.~\cite{lohe}. In this situation, the potential supports minima at $\phi=\pm1$ and $\phi=\pm v$, with classical masses, $m^2_{\pm1}=4(v^2-1)^2$ and $m^2_{\pm v}=4v^2(v^2-1)^2$. Therefore, for $v=1$ we only have two minima with classical masses $m^2_{\pm1}=0$, so the solutions engender power-law tails. For $v=0$, we have three minima, with classical masses $m^2_{\pm1}=4$ and $m^2_{0}=0$, so there is exponential (power-law) falloff in the tail connecting the minimum $\phi=\pm1$ ($\phi=0$). 

Let us consider the potential
\be\label{Va}
V(\phi) = \frac{1}{2a}\left|a+(1-a)\phi^2\right|\left(1-\phi^2\right)^2,
\ee
where $a$ is a positive real parameter. A similar model was considered in Ref.~\cite{tdlee}, but our model allows us to get a fixed maximum and classical mass modified by the parameter $a$; this leads to distinct features. The case $a=1$ recovers the potential in Eq.~\eqref{vphi4}. For $0<a<1$, the potential supports two minima, at $\phi=\pm1$, both of them with classical mass $m^2_{\pm1}=4/a$. The point $\phi=0$ is a local maximum for $1/3\leq a<1$ and minimum for $0<a<1/3$. Interestingly, for $0<a<1/3$, potential presents two symmetric maxima, at $\phi=\pm\sqrt{(3a-1)/(3a-3)}$. If $a>1$, the potential engenders the aforementioned two minima at $\phi=\pm1$ and additional others, at $\phi=\pm\sqrt{a/(a-1)}$, with infinite classical mass. The infiniteness of the classical mass of the latter minimum induces the compactification of its associated tail in the solution, leading to a half-compact kink. The points $\phi=0$ and $\phi=\pm\sqrt{(3a-1)/(3a-3)}$ are local maxima. For $0<a\leq1$, there is only a single topological sector, $\phi\in[-1,1]$. For $a>1$, the presence of the aforementioned four minima allows for the existence of three topological sectors, $\phi\in[-\sqrt{a/(a-1)},-1]$, $\phi\in[-1,1]$ and $\phi\in[1,\sqrt{a/(a-1)}]$. Each topological sector supports kink and antikink solutions connecting the minima which define the sector.

\begin{figure}[t!]
\centering
\includegraphics[width=0.5\linewidth]{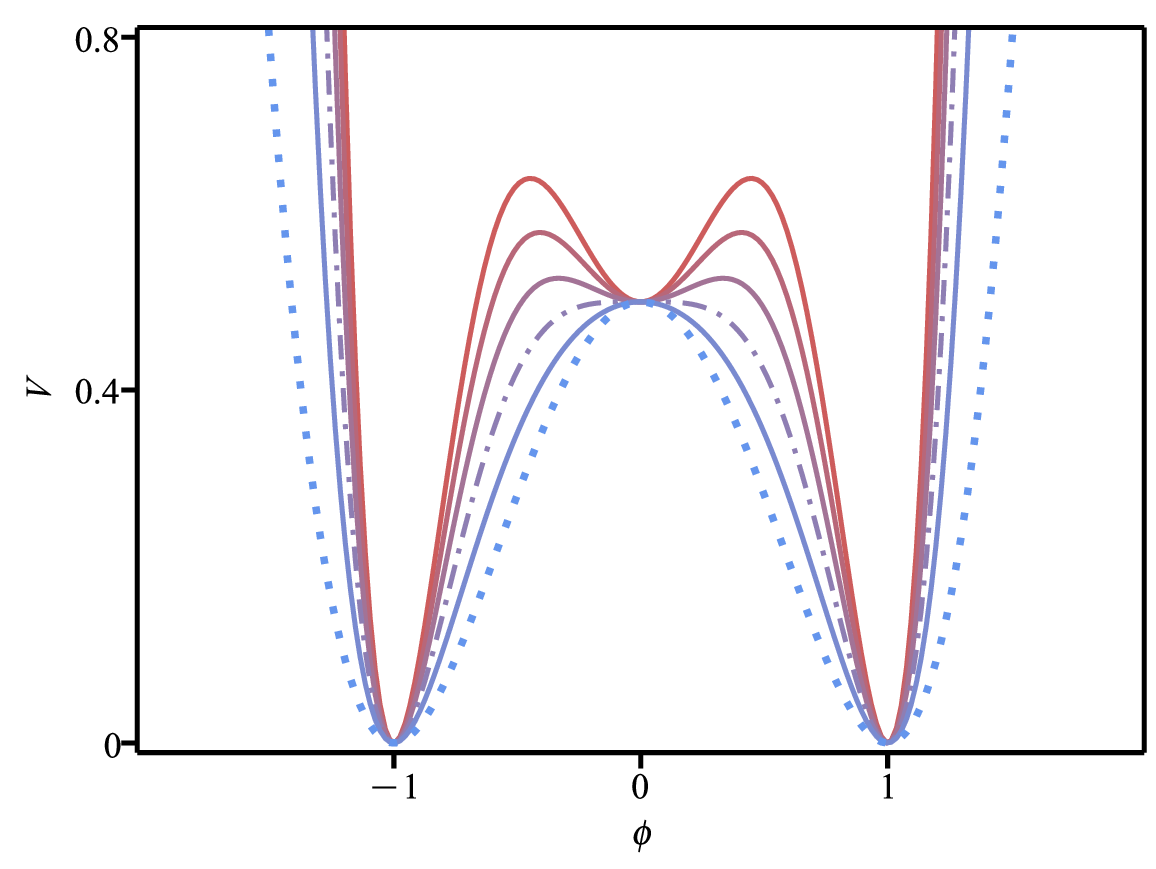}\includegraphics[width=0.5\linewidth]{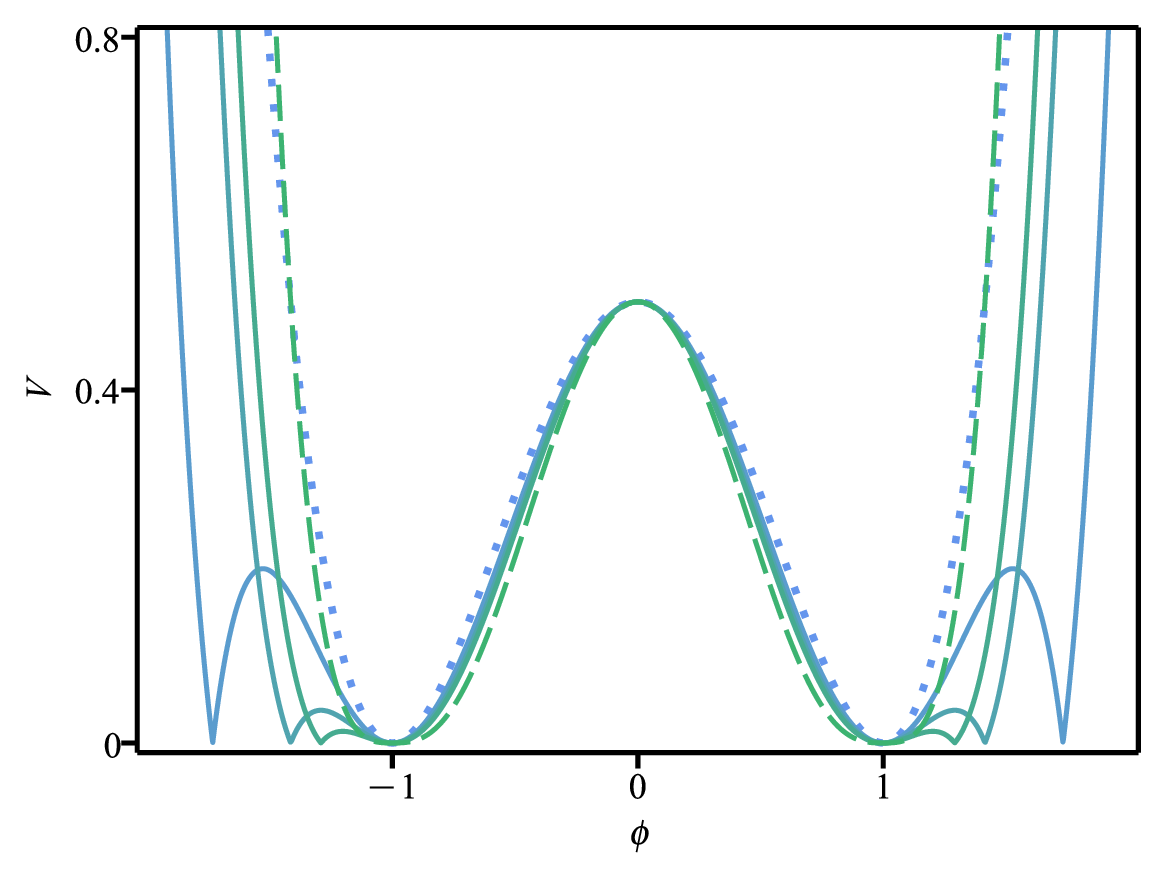}
\caption{The potential $V(\phi)$ in Eq.~\eqref{Va} for $a=1/6,1/5,1/4,1/3,1/2,1$ (left) and $a=1,3/2,2,5/2$ and the limit $a\to\infty$ (right). The dotted lines represent the case $a=1$ as in Eq.~\eqref{vphi4}, the dash-dot line denotes the case $a=1/3$ that delimits the maximum value of $a$ for which the point $\phi=0$ is a maximum, and the dashed line stands for the limit $a\to\infty$ described by Eq.~\eqref{Vinf}.}
\label{fig1}
\end{figure}

\begin{figure}[t!]
\centering
\includegraphics[width=0.7\linewidth]{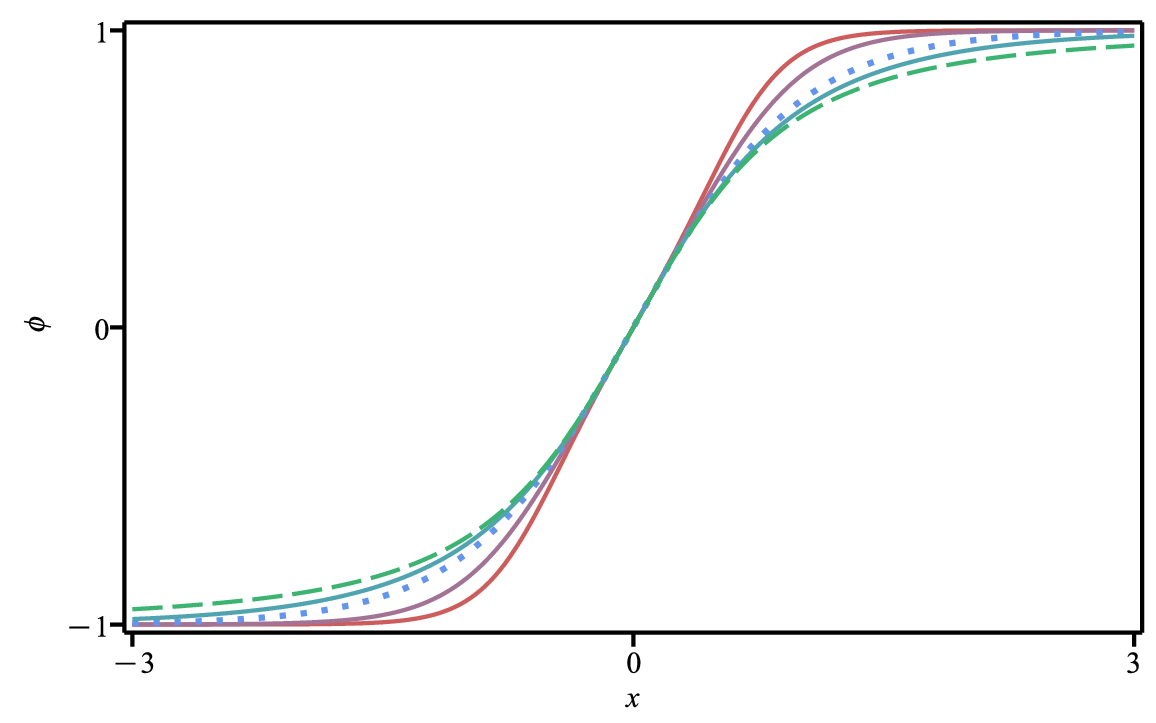}
\caption{Solution $\phi(x)$ in Eq.~\eqref{sola} for $a=1/6,1/3,1,5/2$ and the limit $a\to\infty$ as in Eq.~\eqref{solinf}. The dotted and dashed lines represent, respectively, the case $a=1$ and $a\to\infty$.}
\label{fig2}
\end{figure}

As $a$ gets larger and larger, the classical mass associated to the minima $\phi=\pm1$ tends to zero and the minima with infinite classical mass run away, disappearing for $a\to\infty$. In the limit $a\to\infty$, we have
\be\label{Vinf}
V(\phi) = \frac12\left|1-\phi^2\right|^3,
\ee
whose classical masses is null. This potential was introduced in Ref.~\cite{long1}; it supports solutions with long-range tails. Therefore, we are smoothly going from models that support solutions with exponential to ones with power-law tails. Our procedure, although similar to the one developed in Ref.~\cite{colisao4}, deals with polynomial potentials and, as we shall see, allows for the presence of analytical results. We first consider the solutions in the central sector of the potential \eqref{Va}, $\phi\in[-1,1]$. In Fig.~\ref{fig1}, we depict the potential \eqref{Va} for several values of $a$, including $a=1$ and the limit $a\to\infty$.
%%%%%%%%%%%%%%%%
%%%%%%%%%%%%%%%%
Considering that we are working within the first-order formalism, we can use Eq.~\eqref{fo} to get
\be\label{foa}
\phi' = \left|1-\phi^2\right|\sqrt{\left|1+(a^{-1}-1)\phi^2\right|}.
\ee
The solution that obeys the asymptotic conditions $\phi(\pm\infty)\to\pm1$ is
\be\label{sola}
\phi(x) = \frac{\sqrt{a}\tanh\big(x/\sqrt{a}\big)}{\sqrt{1+(a-1)\tanh^2\!\big(x/\sqrt{a}\big)}},
\ee
where we have taken $\phi(0)=0$. Because $\phi=0$ is a fixed local maximum for any value of $a$, the above function behaves as $\phi(x)\approx x$ near the origin. The asymptotic expression of the solution takes the form $\phi(x)\approx\pm1\mp(2/a)e^{-2|x|/\sqrt{a}}$ for $x\to\pm\infty$. Notice that, as $a$ gets larger and larger, the exponential term tends to become smaller and smaller. Indeed, by taking the limit $a\to\infty$, the solution takes the form
\be\label{solinf}
\phi(x) = \frac{x}{\sqrt{1+x^2}},
\ee
whose asymptotic behavior is $\phi(x)\approx \pm1\mp x^{-2}/2$ for $x$ going to $\pm\infty$. This is compatible with Eq.~\eqref{fluc} for $m^2=4/a$, $\mu=6(2a-1)/a$ and $\alpha=2$. As expected, in the limit $a\to\infty$, we get $m^2=0$, $\mu=12$ and $\alpha=2$, which explains the power-law tail. The above solution was investigated in Ref.~\cite{long1}, where it was called highly interactive due to its long-range tails. We remark that, even though the power-law asymptotic behavior only arises in the limit $a\to\infty$, the solution \eqref{sola} become more and more interactive as $a$ gets larger. This is a consequence of the decreasing of the classical mass as we increase $a$. In Fig.~\ref{fig2}, we display the solution \eqref{sola} for several values of $a$, including the limit $a\to\infty$ in Eq.~\eqref{solinf}.
%%%%%%%%%%%%%%%%
%%%%%%%%%%%%%%%%

The energy density \eqref{rho} associated to the solution \eqref{sola} is
\be\label{rhoa}
\rho(x) = \frac{\sech^4\big(x/\sqrt{a}\big)}{\left(a +(1-a)\,\sech^2\big(x/\sqrt{a}\big)\right)^3}.
\ee
It has a fixed value at the origin, $\rho(0)=1$. This point is a local minimum for $a<1/3$ and is a global maximum for $a\geq1/3$. In the case $a<1/3$, the energy density gets two symmetric maxima around the origin. The limit $a\to\infty$ makes the above expression become
\be\label{rhoinf}
\rho(x) = \frac{1}{\left(1+x^2\right)^3}.
\ee
Therefore, the energy density also supports power-law tails when $a$ tends to infinity. By integrating the energy density \eqref{rhoa}, we get the energy
\be\label{E}
E = \frac{3a-2}{4\sqrt{a}\,(a-1)} +\frac{\sqrt{a}\,(3a-4)M(a)}{4\,|a-1|^{3/2}}.
\ee
In the above expression, $M(a)= -\arctanh\big(\sqrt{1-a}\big)$ for $a<1$, and $M(a)=\arctan\big(\sqrt{a-1}\big)$ otherwise. The above expression can be expanded around $a=1$, in the form
\be
E = \frac43 - \frac{2}{15}(a-1) + {\cal O}\!\left[(a-1)^2\right]
\ee
Therefore, the limit $a\to1$ leads to the energy $E=4/3$, as expected, since $a=1$ recovers the so-called $\phi^4$ model. In the limit $a\to\infty$, we get $E=3\pi/8$. We remark that it diverges in the limit $a\to0$, justifying the exclusion of $a=0$. The energy is monotonically decreasing with $a$.
%%%%%%%%%%%%%%%%
\begin{figure}[t!]
\centering
\includegraphics[width=0.5\linewidth]{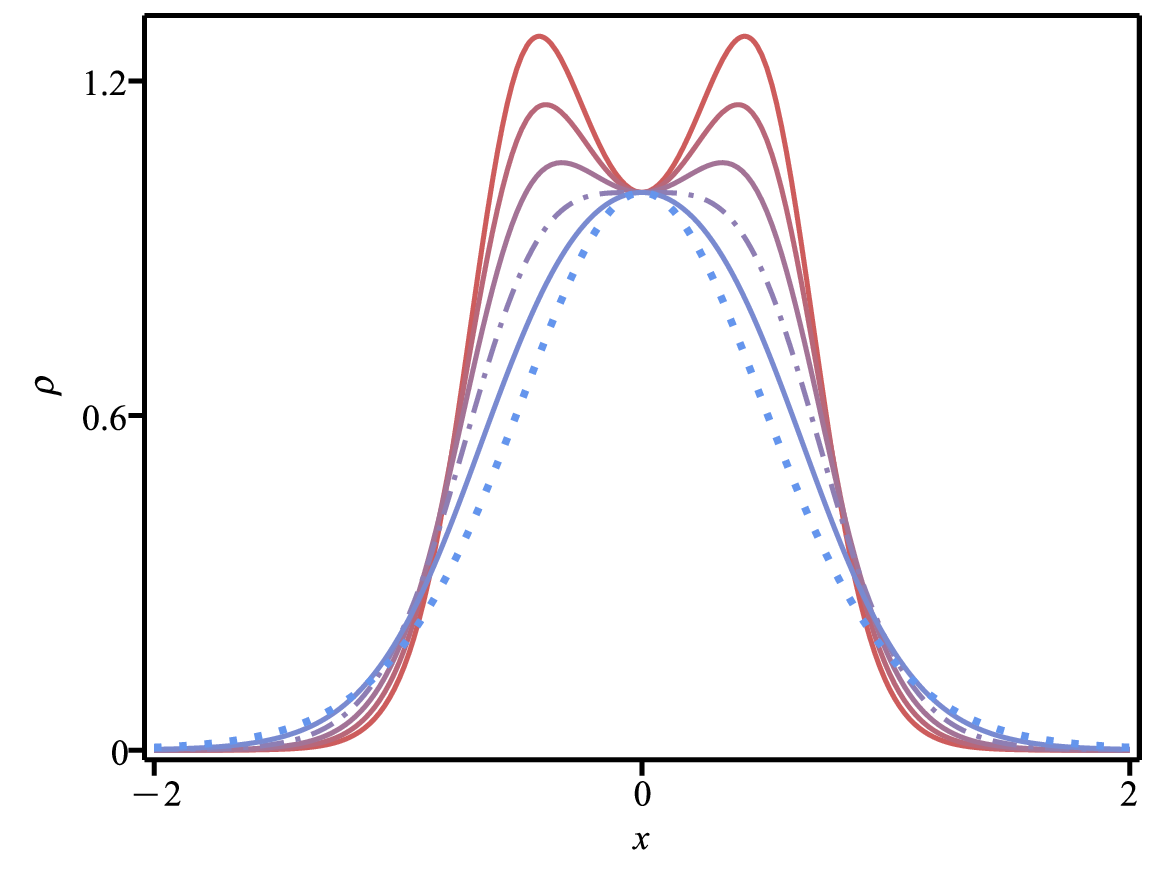}\includegraphics[width=0.5\linewidth]{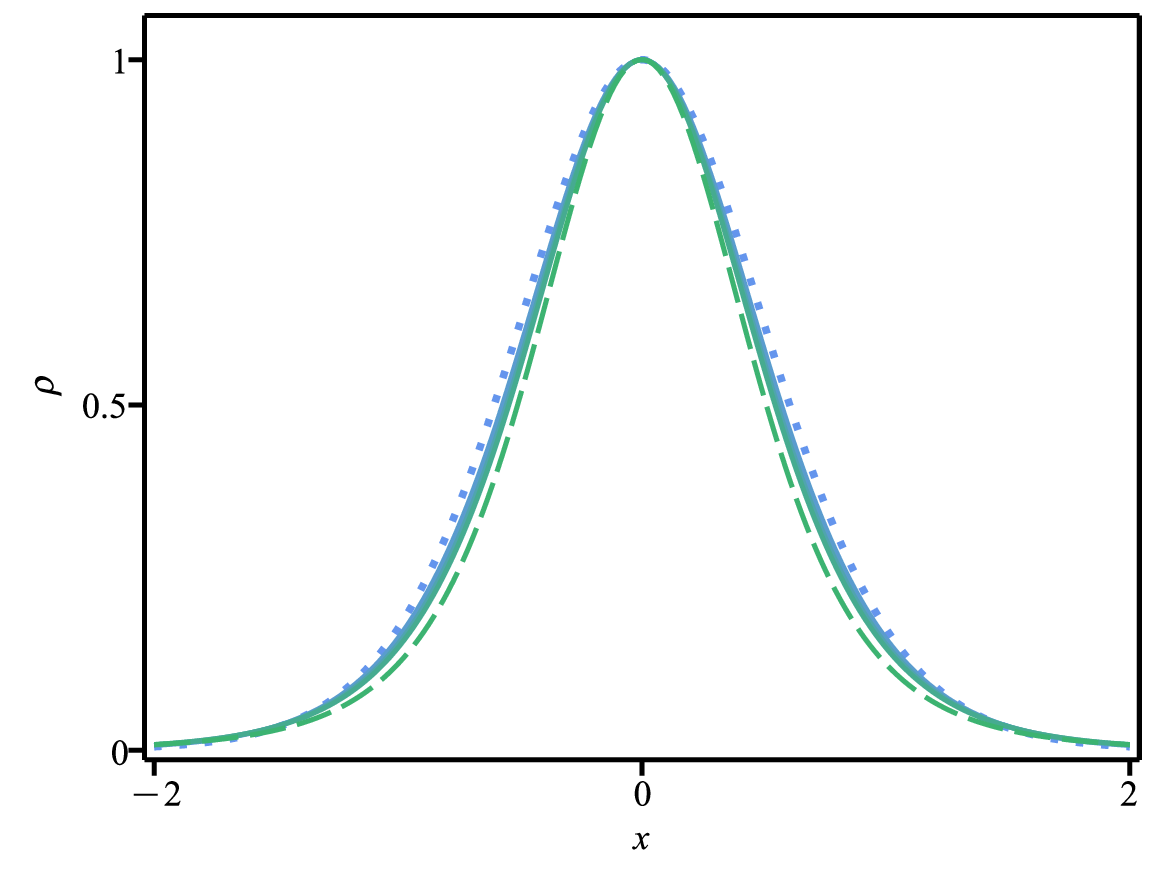}
\caption{The energy density $\rho(x)$ in Eq.~\eqref{rho} for $a=1/6,1/5,1/4,1/3,1/2,1$ (left) and $a=1,3/2,2,5/2$ and the limit $a\to\infty$ (right). The dotted lines represent the case $a=1$, the dash-dot line denotes the case $a=1/3$ that delimits the maximum value of $a$ for which the point $x=0$ is a maximum, and the dashed line stands for the limit $a\to\infty$ described by Eq.~\eqref{rhoinf}.}
\label{fig3}
\end{figure}
%%%%%%%%%%%%%%%%

\begin{figure}
    \centering
\includegraphics[width=0.5\linewidth]{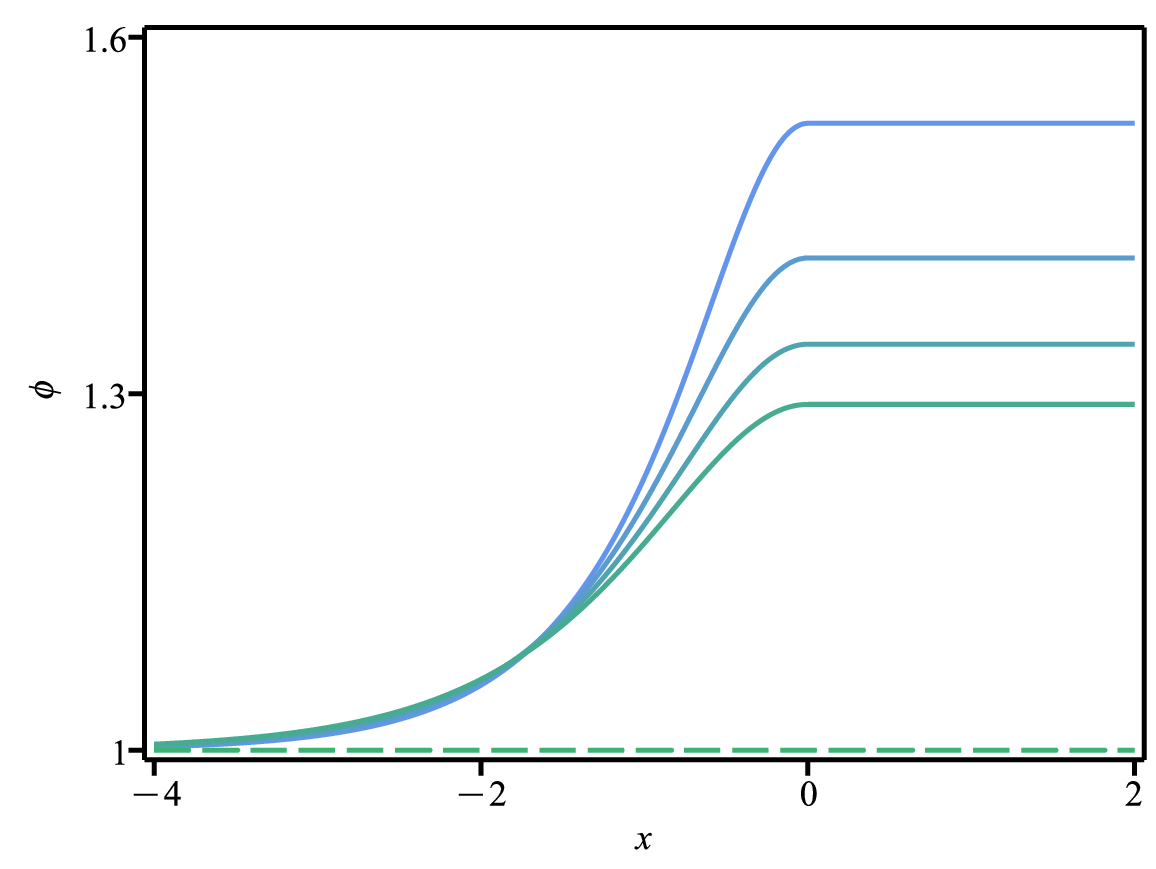}\includegraphics[width=0.5\linewidth]{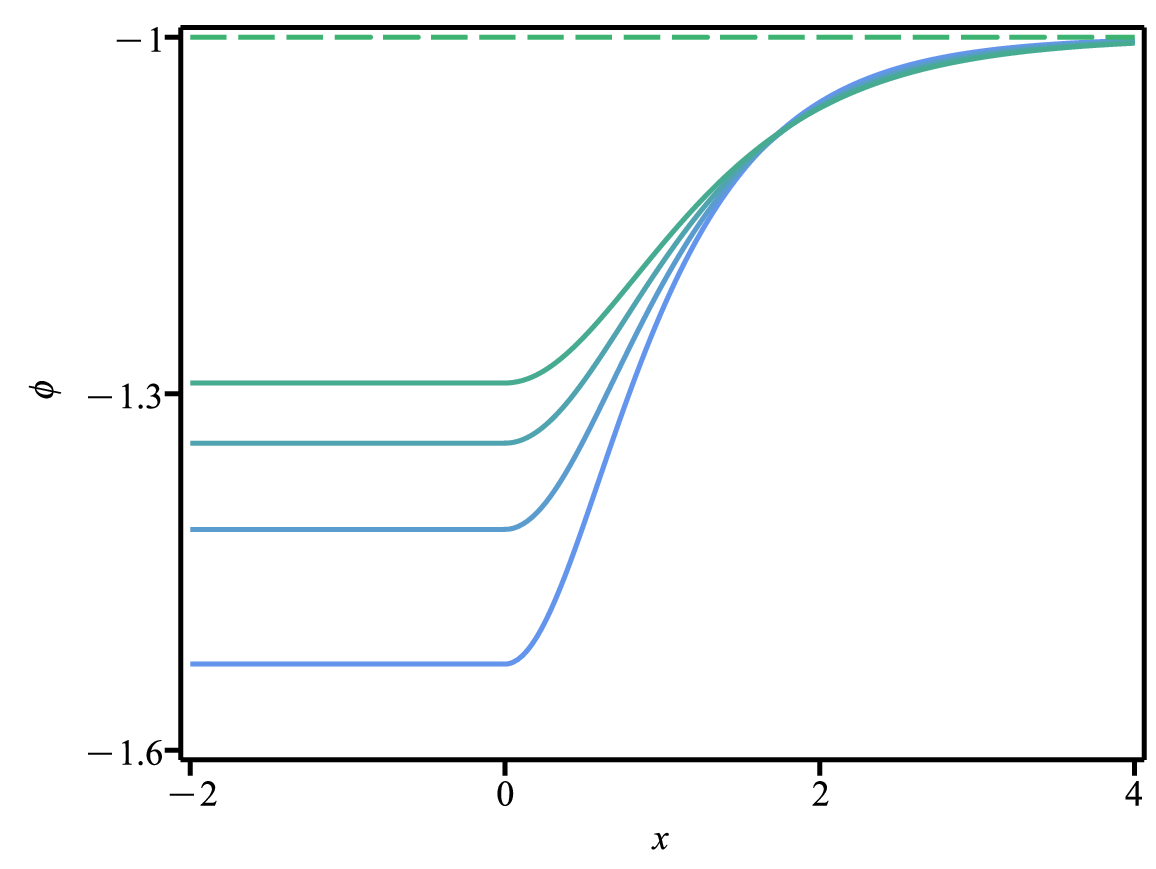}
\includegraphics[width=0.5\linewidth]{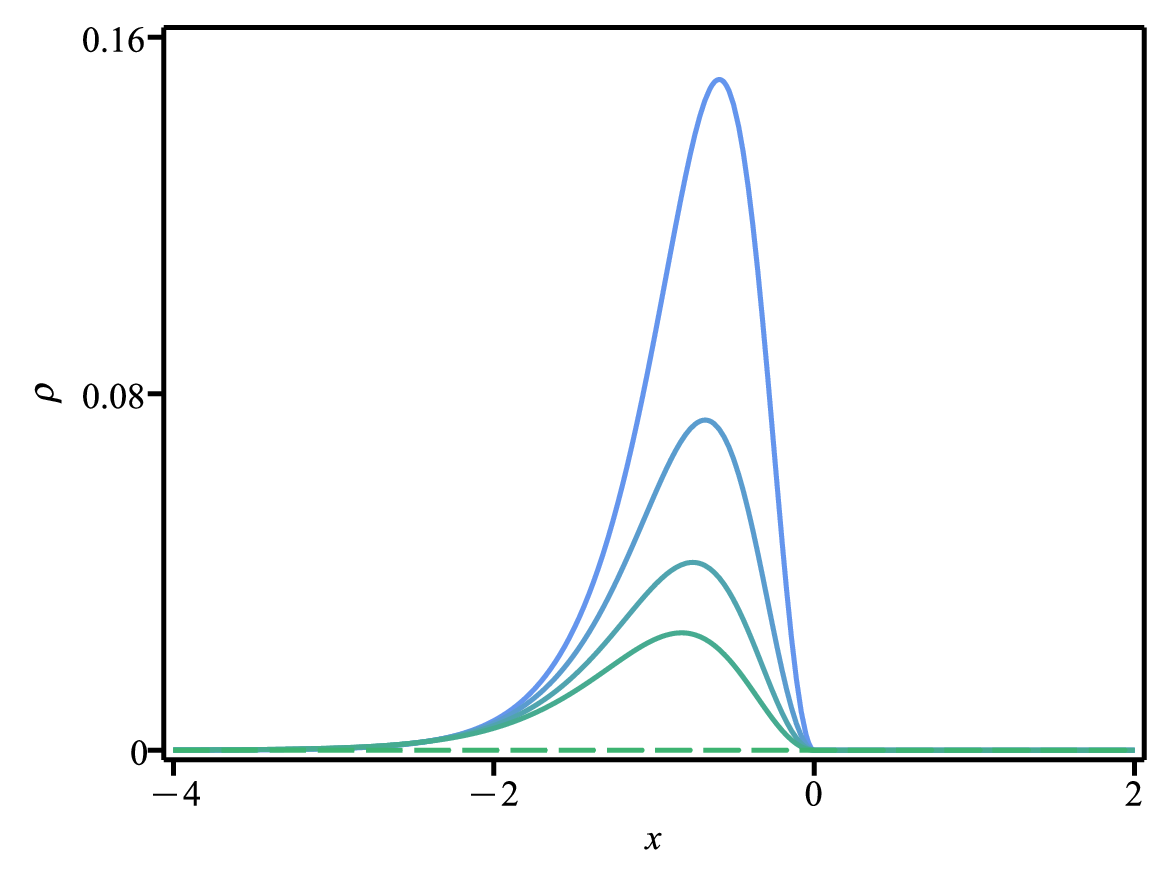}\includegraphics[width=0.5\linewidth]{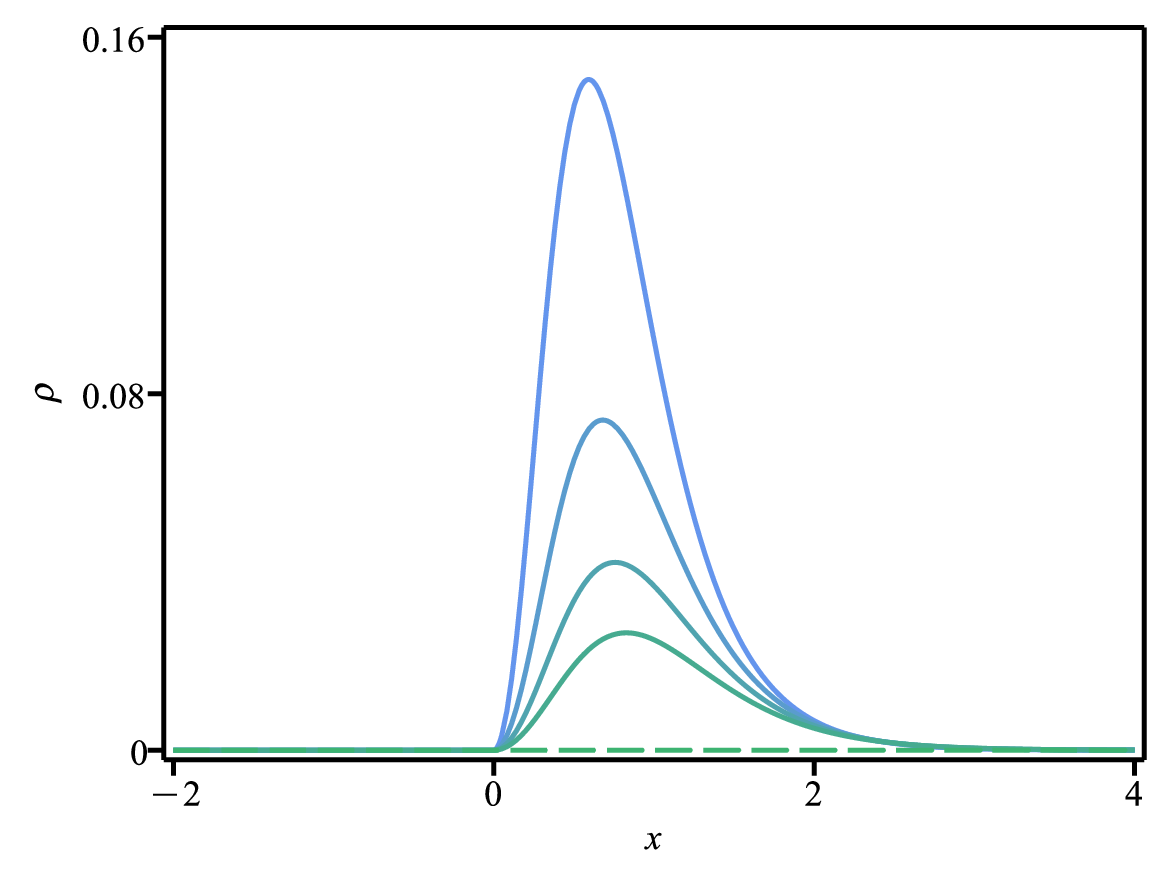}
    \caption{The solution $\phi_+(x)$ (top left) and $\phi_-(x)$ (top right) in Eq.~\eqref{solhca} and the corresponding energy density $\rho_+(x)$ (bottom left) and $\rho_-(x)$ (bottom right) in Eq.~\eqref{rhohca} for  $a=1.75,2,2.25$ and $2.5$. The colors go from blue to green as $a$ increases. The dashed lines represent these quantities in the limit $a\to\infty$, which leads to uniform solutions ($\phi_+(x)=1$ and $\phi_-(x)=-1$) and null energy density.}
    \label{fig4}
\end{figure}

The expressions \eqref{sola}--\eqref{E} describes the interval $\phi\in[-1,1]$ for $a>0$. However, as we have previously commented, the potential supports minima outside the aforementioned interval if $a>1$. In this situation, the first-order equation \eqref{foa} can be solved for the external sectors of the potential \eqref{Va}, $\phi\in I_-\equiv\big[-\sqrt{a/(a-1)},-1\big]$ and $\phi\in I_+\equiv\big[1,\sqrt{a/(a-1)}\big]$. We then get
\be\label{solhca}
\phi_\pm(x) = 
\begin{cases}
-\cfrac{\sqrt{a}\coth\big(x/\sqrt{a}\big)}{\sqrt{1+(a-1)\coth^2\big(x/\sqrt{a}\big)}} \quad&\text{for $\pm x<0$}\\
\pm\sqrt{\cfrac{a}{a-1}} \quad&\text{for $\pm x\geq0$},
\end{cases}
\ee
where the $\phi_\pm(x)$ stands for the solution in the sector $I_\pm$. Notice that the solutions are now half-compact: the tails associated to the minima $\phi=\pm\sqrt{a/(a-1)}$ are compact, whilst the ones going to the minima $\phi=\pm1$ are exponential. Interestingly, the extension of the tail which exponential falloff depends on $a$, according to $\phi_\pm(x) \approx \pm1 \pm(2/a)e^{-2|x|/\sqrt{a}}$ for $x\to\pm\infty$. Moreover, the parameter $a$ controls the amplitude of the solutions: as it gets larger, the amplitude gets smaller. In the limit $a\to\infty$, the above solutions become uniform, $\phi_\pm=\pm1$. From Eq.~\eqref{rho}, we get the energy density
\be\label{rhohca}
\rho_\pm(x) = 
\begin{cases}
\cfrac{\tanh^2\big(x/\sqrt{a}\big)\,\sech^4\big(x/\sqrt{a}\big)}{\left(a-\sech^2\big(x/\sqrt{a}\big)\right)^3} \quad&\text{for $\pm x< 0$}\\
0 \quad&\text{for $\pm x\geq 0$}.
\end{cases}
\ee
The parameter $a$ controls the maximum of the energy density, given by $\rho_\pm(x^*_\pm)=2\sqrt{3}/(9\sqrt{a}(a-1))$, with $x_\pm^* = \pm\sqrt{a}\arccos\big(\sqrt{(3a-1)/(2a)}\big)$. As $a$ increases, the height of the energy density goes down. In the limit $a\to\infty$, the energy density is null, as expected, because the solution is uniform. The solutions \eqref{solhca} and the above energy density can be seen in Fig.~\ref{fig4}.
%%%%%%%%%%%%%%%%

Interestingly, both half-compact solutions have the same energy, calculated by the integration of the above expression in all the space,
\be\label{Ehc}
E = \frac{3a-2}{8\sqrt{a}\,(a-1)} -\frac{\sqrt{a}\,(3a-4)\big(\pi-2\arctan\big(\sqrt{a-1}\big)\big)}{16\,(a-1)^{3/2}}.
\ee
This energy is monotonically decreasing with $a$. It vanishes in the limit $a\to\infty$, as expected, since it leads to uniform solutions. 

\section{Linear Stability}
Since we are dealing with a model with a parameter that modifies the tail of the solution, let us now investigate its stability against small fluctuations. By taking $\phi(x,t)=\phi_s(x)+\psi(x,t)$ in Eq.~\eqref{eom}, where $\phi_s(x)$ is the solution of Eq.~\eqref{fo} and $\psi(x,t)$ is the perturbation, we get
\be
\Box\psi +U(x)\psi = 0,
\ee
in which $U(x)=V_{\phi\phi}\big|_{\phi=\phi_s(x)}$ is the stability potential. The above expression allows that we separate space and time coordinates in the form $\psi(x,t)=\sum_n\psi_n(x)\cos(\omega_nt)$, such that the stability is now governed by
\be
-\psi_n^{\prime\prime} +U(x)\psi_n = \omega_n^2\psi_n.
\ee
This is a one-dimensional Schr\"odinger-like eigenvalue equation. Notice that the stability potential tends asymptotically to the square classical masses \eqref{classicalmass} associated to the minima connected by the solution $\phi_s(x)$, i.e., $U(\pm\infty)=m^2_{v_i}$. We can follow the lines of Refs.~\cite{modozero1,modozero2} to show that the zero mode $\omega_0=0$ exists; it has the form
\be\label{psi0}
\psi_0(x) = \frac{\phi^\prime}{\sqrt{E}},
\ee
where $E$ is the energy of the solution.

Our model has two types of solutions. Let us first consider the one in the central sector of the potential \eqref{Va}, described by \eqref{sola}. In this situation, the stability potential is
\be\label{Ua}
U(x) = \frac{4a^2+2a(2a-5)\,S^2-(8a+1)(a-1)\,S^4}{a\left(a+(1-a)\,S^2\right)^2},
\ee
where $S=\sech\big(x/\sqrt{a}\big)$. In the limit $|x|\to\infty$, the stability potential is $U(\pm\infty)=4/a$. At the origin, we have $U(0)=(1-3a)/a$; this point is a global minimum for $a\geq2/3$ and a local maximum for $a<2/3$, surrounded by two symmetric minima. For $a=1/3$, in particular, we have $U(0)=0$. 

\begin{figure}[t!]
\centering
\includegraphics[width=0.5\linewidth]{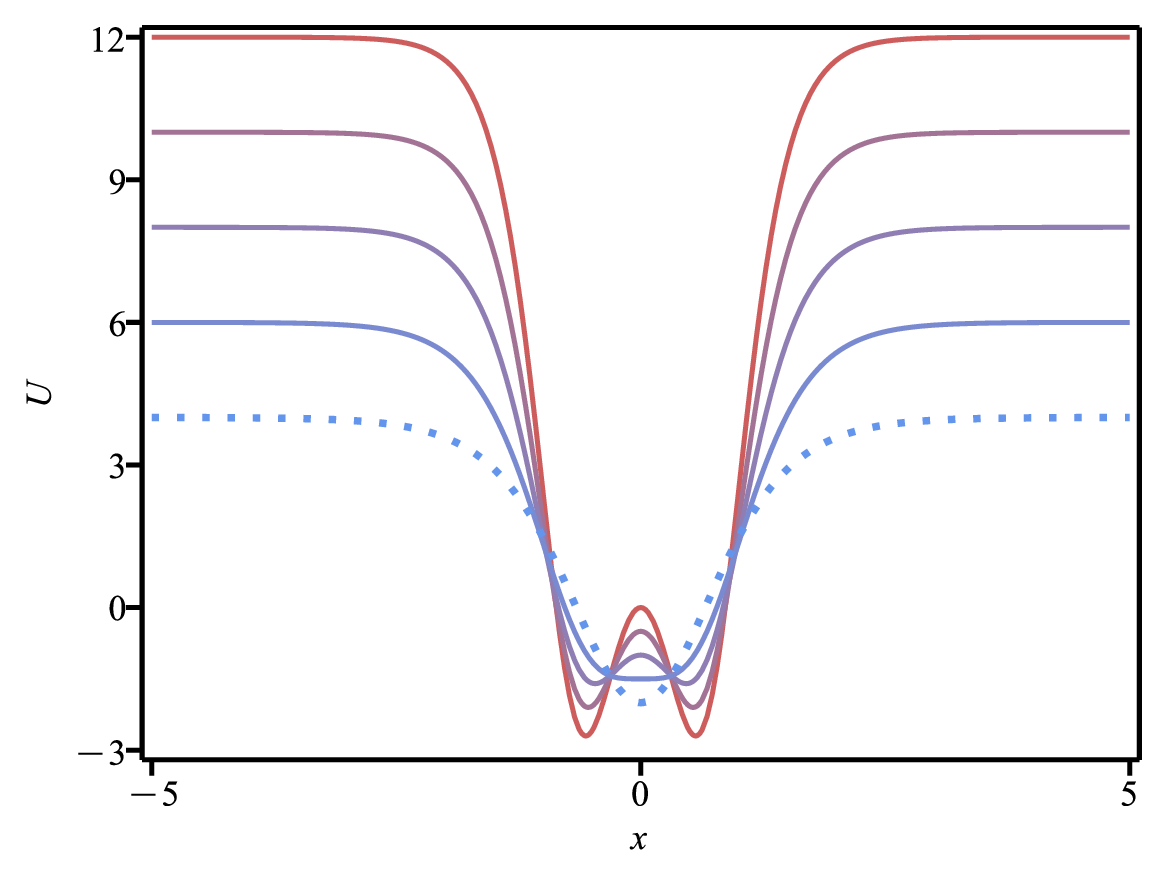}\includegraphics[width=0.5\linewidth]{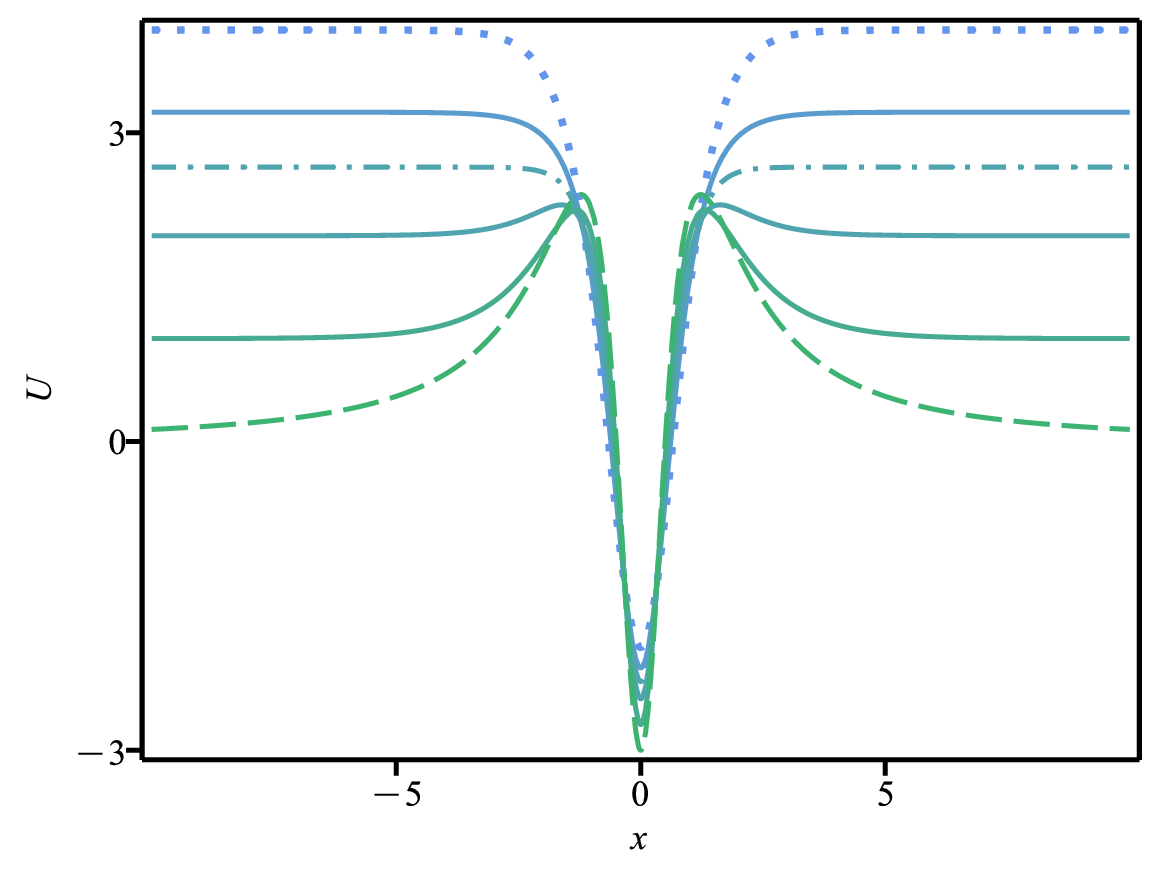}
\caption{The stability potential $U(x)$ in Eqs.~\eqref{Ua}, for $a=1/3,2/5,1/2,2/3$ and $1$ (left) and $a=1,5/4,3/2,2,4$ and $a\to\infty$ (right). The dotted lines represent the case $a=1$, the dash-dot line denotes the case $a=3/2$ that delimits the maximum value of $a$ for which the stability potential does not engender a volcano profile, and the dashed line stands for the limit $a\to\infty$ described by Eq.~\eqref{Uinf}. In the left panel, the curve with lowest asymptotic value is represented by $a=1$ (dotted line); as $U(\pm\infty)$ goes up, $a$ decreases. In the right panel, $a=1$ (dotted line) stands for the highest asymptotic value; as $U(\pm\infty)$ goes down, $a$ increases.}
\label{fig5}
\end{figure}

Interestingly, the stability potential has its asymptotic value ($U(\pm\infty)=4/a$) as its maximum value for $a\leq3/2$; this means that semi-bound states are present. This behavior changes if $a>3/2$, for which it has a volcano shape, with the asymptotic values becoming smaller than its maxima, so there are no semi-bound states. However, there is a bound state at $\omega^2=4/a$ for $a\leq1.577$. In the limit $a\to\infty$, the above stability potential becomes
\be\label{Uinf}
U(x) = \frac{12x^2-3}{\left(1+x^2\right)^2},
\ee
which has a volcano profile that vanishes asymptotically. We display, in Fig.~\ref{fig5}, the stability potential \eqref{Ua} for some values of $a$, including the limit $a\to\infty$ given above.
%%%%%%%%%%%%%%%%
%%%%%%%%%%%%%%%%
The zero mode \eqref{psi0} associated to the solution \eqref{sola} is
\be
\psi_0(x) = \frac{S^2}{\sqrt{E}\left(a+(1-a)\,S^2\right)^{3/2}}.
\ee
In the limit $a\to\infty$, it can be written as
\be
\psi_0(x) = \frac{2\sqrt{6}}{3\sqrt{\pi}\left(1+x^2\right)^{3/2}}.
\ee
For all values of $a$, the zero mode does not engender nodes. This ensures that it is the ground state and there is no negative eigenvalues. Therefore, the solution \eqref{sola} is stable under small fluctuations.

\begin{figure}[t!]
\centering
\includegraphics[width=0.5\linewidth]{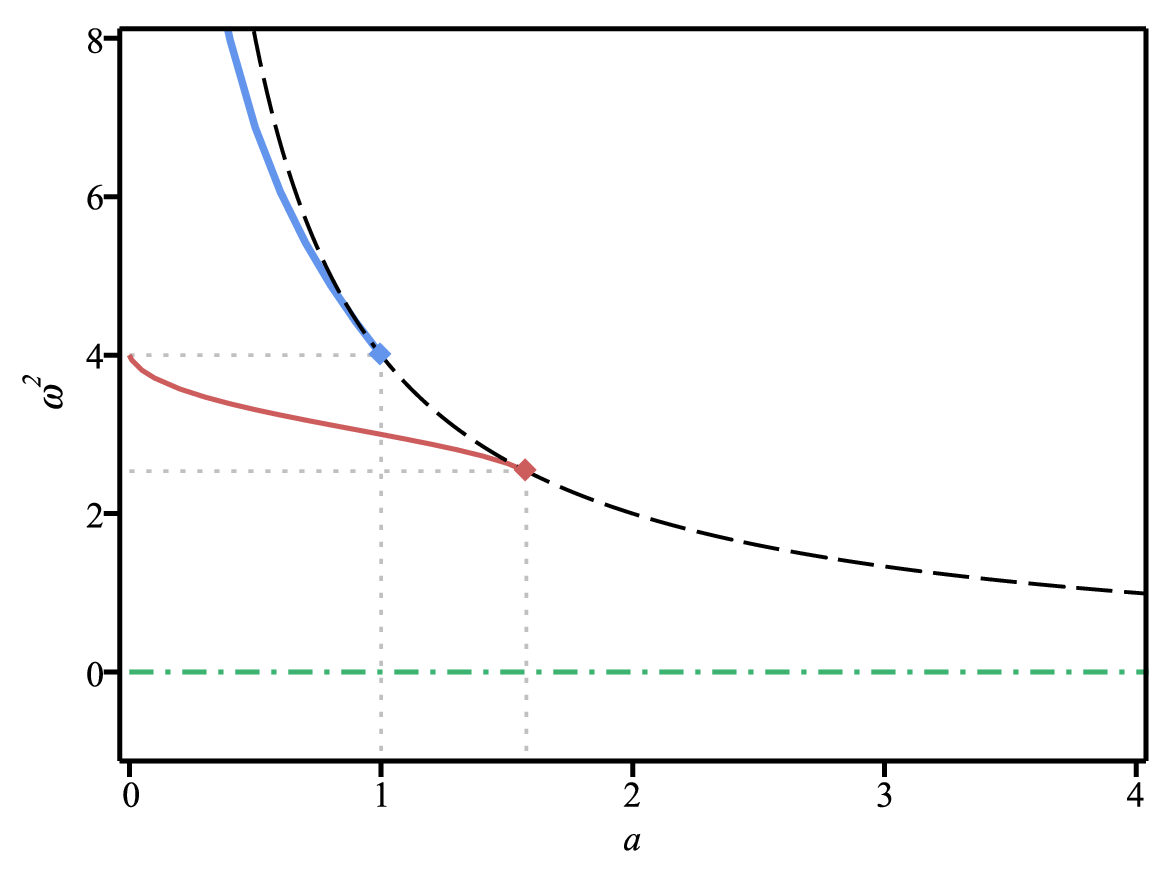}\includegraphics[width=0.5\linewidth]{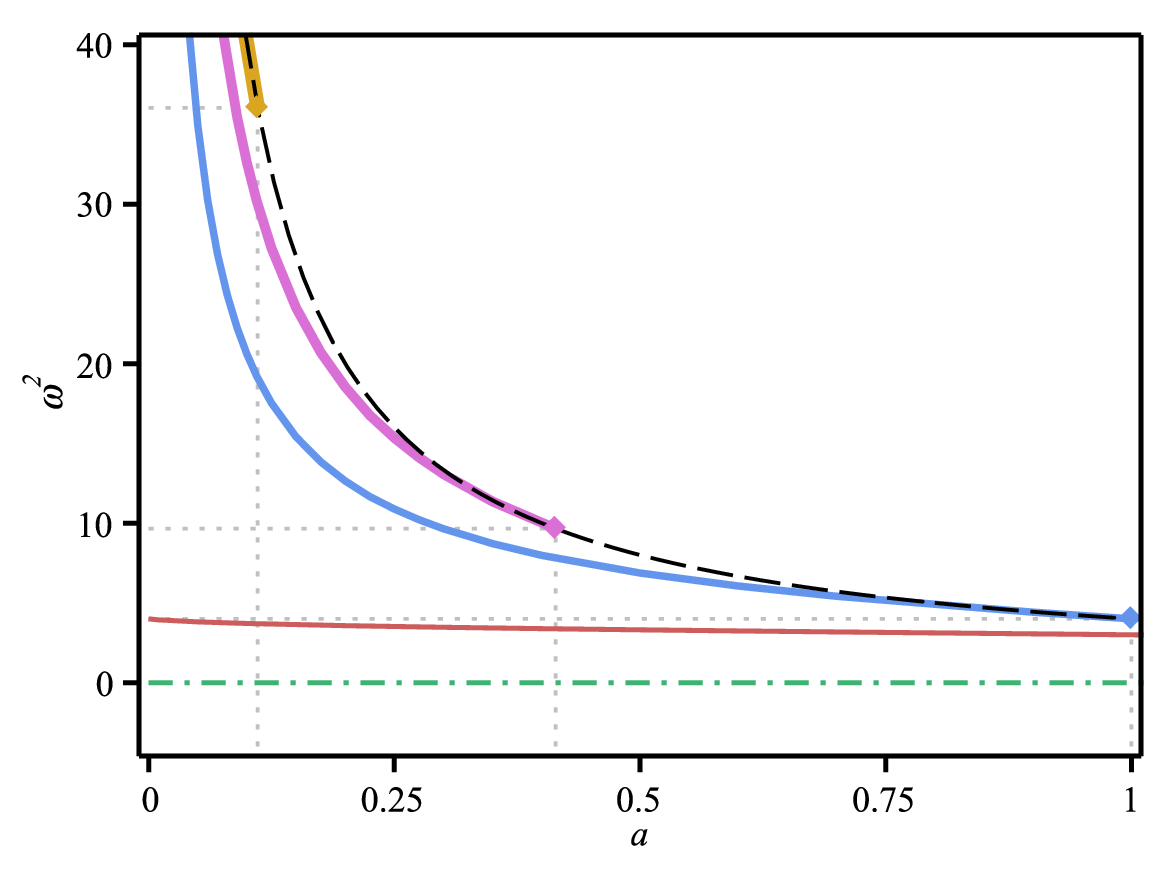}
\caption{The eigenvalues $\omega^2$ in terms of the parameter $a$ in the range $(0,4]$ (left) and $(0,1]$ (right). The dashed lines represent the classical mass, $m^2_\pm=U(\pm\infty)$, associated to the stability potential in Eq.~\eqref{Ua}. The green dash-dotted lines represent the zero mode. The red, blue, purple and orange lines denote, respectively, the first, second, third and fourth excited states. The thickness of the solid lines increases as one ranges from the first to the fourth excited state. The diamonds represent the semi-bound states.}
\label{fig6}
\end{figure}

The eigenvalues and number of bound states depend on the parameter $a$. To find them, one must use numerical procedures. For $a>1.577$, we only have a single bound state, which is the zero mode. In the range $1.577\geq a>1$, we have two bound states. In particular, for $a=1.577$, we have $\omega_1^2=2.536$ in addition to the zero mode, at the same height of the asymptotic values of the stability potential. For $a=1$, there are two bound state and a semi-bound state at the asymptotic value of $U(x)$, $\omega^2=4$ \cite{modozero1}. In the interval $1>a>0.414$, there are three bound states. The case $a=0.414$ has two bound states and a semi-bound state at $\omega^2=9.661$. Four bound states arise if $0.414>a>0.111$ and at least five bound states are present for $a<0.111$. \emph{In summa}, as $a$ approaches zero, more and more bound states appear. In Fig.~\ref{fig6}, we display the behavior of the bound states with $a$. It is worth commenting that the presence of the vibrational modes may lead to resonance phenomena in kink-antikink collisions. This was shown in Ref.~\cite{gani3}, where the authors studied collisions of kinks with power-law tails and found that a rich resonance phenomenology.
%%%%%%%%%%%%%%%%
%%%%%%%%%%%%%%%%

We must also analyze the stability of the half-compact solutions \eqref{solhca} associated to the external sectors of the potential \eqref{Va}. The stability potential can now be written as
\be\label{uhc}
U_\pm(x) = 
\begin{cases}
\frac{4a^2-2a(6a-5)\,S^2-(3a-1)(a-1)\,S^4}{a\left(a-S^2\right)^2}, \quad&\text{for $\pm x<0$}\\
\infty, \quad&\text{for $\pm x\geq 0$}.
\end{cases}
\ee
The zero mode is calculated from Eq.~\eqref{psi0}; it has the form
\be
\psi^\pm_0(x) = 
\begin{cases}
\cfrac{\tanh\big(x/\sqrt{a}\big)\,\sech^2\big(x/\sqrt{a}\big)}{\sqrt{E}\left(a-\sech^2\big(x/\sqrt{a}\big)\right)^{3/2}} \quad&\text{for $\pm x< 0$}\\
0 \quad&\text{for $\pm x\geq 0$}.
\end{cases}
\ee
This zero mode does not present nodes, ensuring the stability of the solution. Moreover, by analyzing the eigenvalue equation, we have found that $\psi^\pm_0(x)$ is the only bound state.

It is worth commenting that, even though we have shown that the solution is stable under scalar perturbations, there are other stabilities that may be of interest. For instance, one may investigate the stability against contractions and dilations. In this direction, as shown in Ref.~\cite{trilogia1}, the presence of the first-order equation \eqref{foa} is compatible with the stressless condition. This ensures that the solution also comply with this stability criterion. Moreover, the non-negative character of the potential \eqref{Va} and the topological character of the solution \eqref{sola} ensures that the Bogomol'nyi bound \cite{bogo} is saturated, so the energy of the solution \eqref{sola} is a global minimum of the system. Therefore, the solution cannot decay, being stable.

%%%%%%%%%%%%%%%%
\begin{figure}[t!]
\centering
\includegraphics[width=0.5\linewidth]{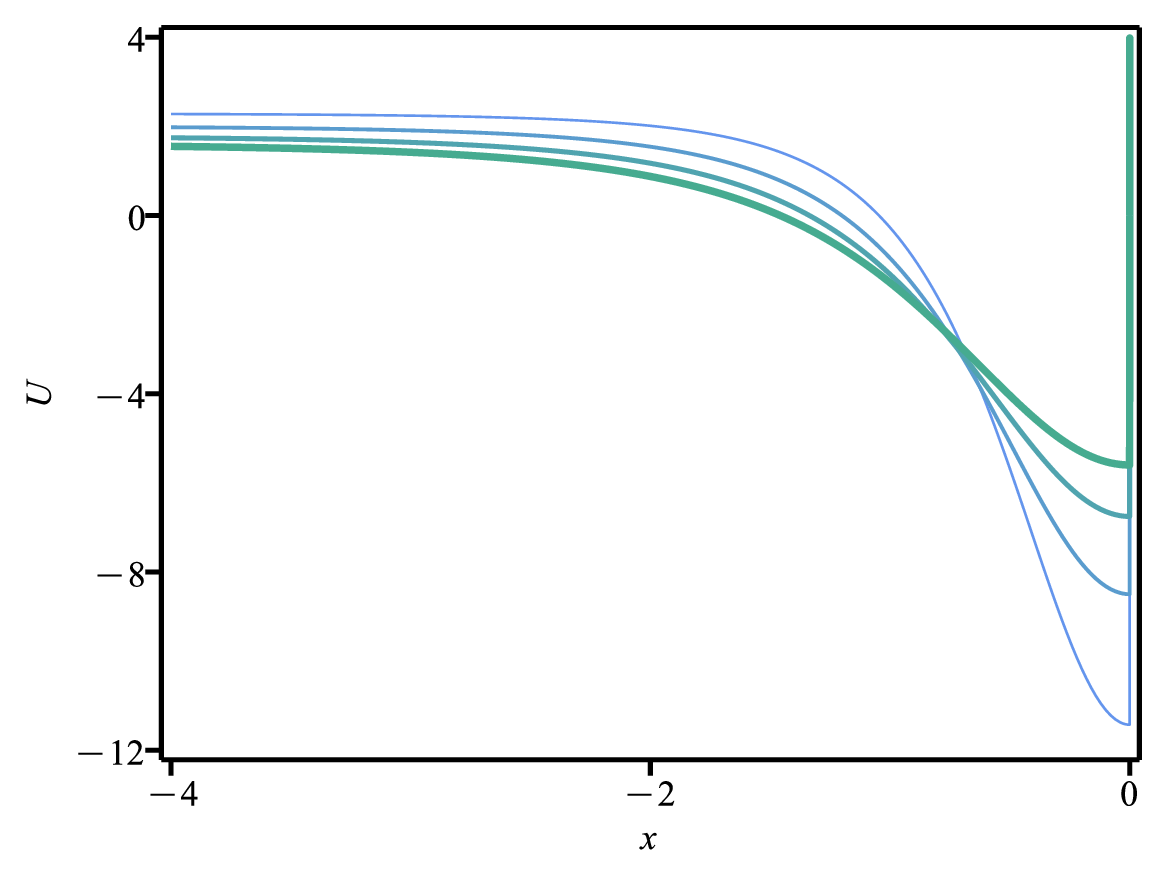}\includegraphics[width=0.5\linewidth]{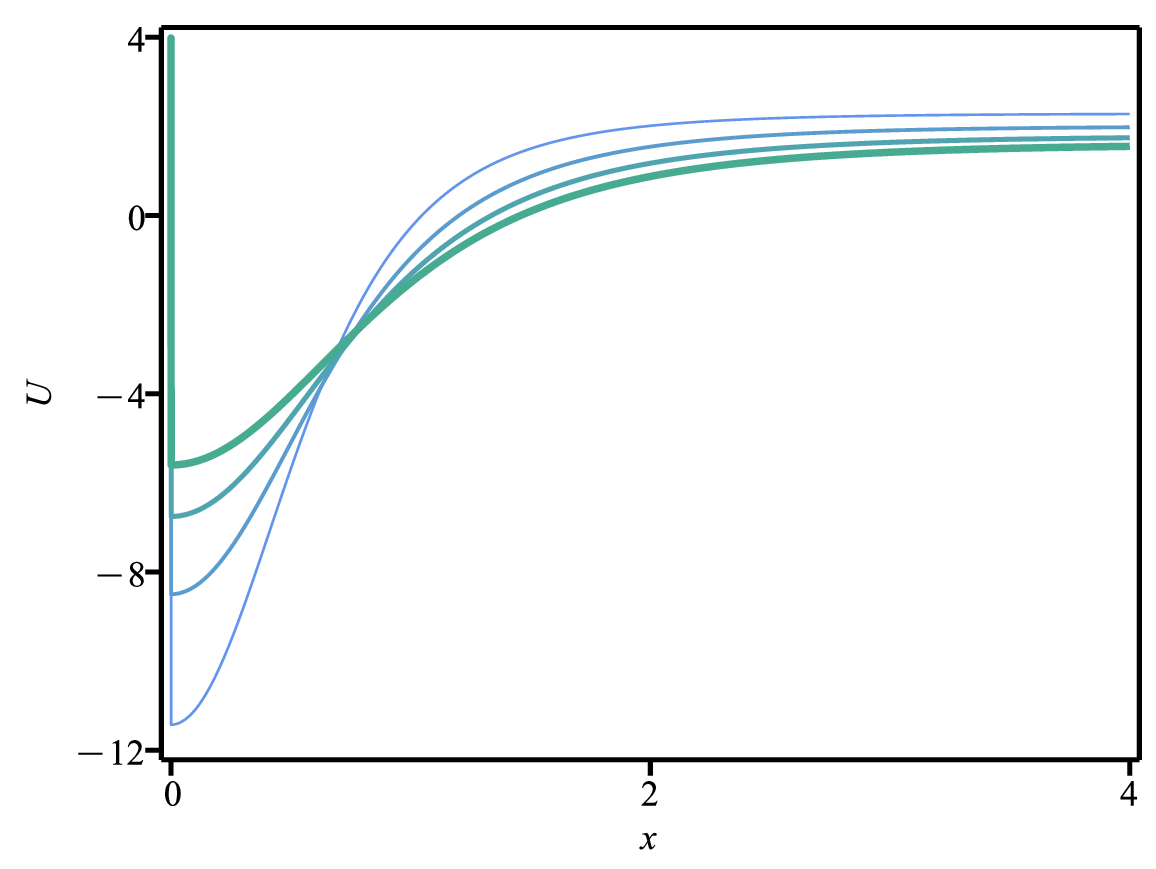}
\caption{The finite region of the stability potential $U_+(x)$ (left) and $U_-(x)$ (right) in Eq.~\eqref{uhc} for $a=1.75,2,2.25$ and $2.5$. The thickness of the lines increases with $a$.}
\label{fig7}
\end{figure}
%%%%%%%%%%%%%%%%

\section{Outlook}
In this paper, we have investigated a class of polynomial potentials that allows us to increase the range of kinks with analytical results. The procedure takes advantage of the first-order framework developed by Bogomol'nyi \cite{bogo} and consists of introducing a parameter in the potential that controls the classical masses.

First, we have considered the central sector $\phi\in[-1,1]$ of the potential \eqref{Va} to show that, as the parameter $a$ gets larger, the solution gets a longer range, going from exponential to power-law tails. This transition also appears in the energy density. We have also investigated the linear stability to show that the parameter $a$ also modifies the shape of the stability potential, that tends to become a volcano as $a$ gets larger; this impacts the eigenvalues and bound states associated to the Schr\"odinger-like equation. Previously, in Ref.~\cite{colisao4,sineduplobazeia}, a similar feature was investigated in a double sine-Gordon model that contains a parameter that plays the same role of $a$. Recently, another mechanism to go from short- to long-range kinks was proposed in a two-field model \cite{2fieldsasympt}, in which the modification in the tail is induced by one of the fields. In our model, described by the potential \eqref{Va}, the transition from exponential to power-law tails occurs exclusively due to the self-interaction of the field, without extra degrees of freedom.

Since the potential engenders two external minima for $a>1$, we have also studied the behavior of the solution in the corresponding external sectors. In this situation, the solutions have a tail similar to the previous case at one side and a compact profile at the other side, so we have called it half-compact. We have shown that, even though the extended tail engenders an exponential falloff, the range of the tail increases with $a$. This shows that this mechanism works even in the presence of such atypical solutions. The half-compact profile is also present in the energy density and leads to an infinite barrier in the stability potential.

We remark that this procedure can be extended to other polynomial potentials. For instance, one may investigate
\be\label{new1}
V(\phi) = \frac12\left(1+2b\phi+b\phi^2\right)\left(1-\phi^2\right)^2,
\ee
with $b\in[0,1]$. This potential engender minima at $\phi=\pm1$, with classical mass $m_{-1}^2=4(1-b)$ and $m_{+1}^2=4(3b+1)$. In this situation, $b$ works differently from the corresponding parameter in \eqref{Va}: as $b$ approaches $1$, the range of the solutions increase without changing the location of the minima of the potential. Notice that the behavior of the classical masses is different with $b$: while the range of one tail of the solution increases, the other one decreases with $b$. In this sense, $b$ induces an asymmetry in the kink-like configuration. The specific situation in which $b=1$ was investigated in Ref.~\cite{longrangephi6}. The general model with the above potential requires further study, but we will not deal with it any further here because it does not provide analytical solutions. 

If one wishes to get symmetric long-range configuration, one may consider the potential
\be\label{new2}
V(\phi) = \frac12\!\left(1\!+\!2b\phi\!+\!b\phi^2\right)\left(1\!-\!2b\phi\!+\!b\phi^2\right)\!\left(1-\phi^2\right)^2\!,\!\!\!
\ee
with $b\in[0,1]$. It supports minima at $\phi=\pm1$ with classical mass $m_{\pm1}^2=4(3b+1)(1-b)$. Therefore, for $b=1$, the solutions get power-law tails, engendering long-range behavior. Although we cannot get analytical solutions, we can conclude that they are symmetric because of the symmetry of the corresponding topological sector of this new potential \eqref{new2}.

As perspectives, we would like to suggest that, since the solutions \eqref{sola} and \eqref{solhca} are analytical, one may consider the study of their interactions and collisions, in line with the recent works \cite{colisaobazeiahig,forcelongimp,colisaohalfcomp}. The long-range profile described by power-law tails appears only when the minima of the external sector glue into the ones of the central sector of the potential. This may lead to interesting results, revealing the influence of the external minima in the collision process.

\acknowledgments{We would like to thank Dionisio Bazeia for discussions. We also acknowledge financial support from the Brazilian agencies Conselho Nacional de Desenvolvimento Cient\'ifico e Tecnol\'ogico (CNPq), grants Nos. 402830/2023-7 (MAM and RM), 306151/2022-7 (MAM) and 310994/2021-7 (RM), and Paraiba State Research Foundation (FAPESQ-PB) grant No. 2783/2023 (IA).}

%\appendix

%\bibliography{biblio} 

\end{document}